\documentclass[prb,aps,twocolumn,10pt,floatfix,nofootinbib,superscriptaddress]{revtex4-2}

\usepackage{amsmath}
\usepackage{amssymb}
\usepackage{mathtools}

\usepackage{amsthm}
\usepackage{stackrel}
\usepackage[normalem]{ulem}
\usepackage[flushleft]{threeparttable}

\usepackage{amsfonts}
\usepackage{txfonts}
\usepackage{dsfont}   
\usepackage{pifont}                % double stroke font
\usepackage{bbold}                  % \mathbb{0}

\usepackage{array}
\usepackage{dcolumn}                % cell alignment at decimal point
\usepackage{makecell}               % linebreak in table cell
\usepackage{multirow}

\usepackage{enumerate}
\usepackage[shortlabels]{enumitem}

\usepackage[usenames,dvipsnames]{xcolor}
\usepackage{graphicx}
\graphicspath{{figs/}} % Setting the graphics path

\usepackage[version=4]{mhchem}      % chemical formulae
\usepackage{physics}                % physics notation
\usepackage{csquotes}               % quotations

\usepackage[colorlinks=true,citecolor=cyan,linkcolor=magenta,filecolor=magenta]{hyperref}
\usepackage[capitalize]{cleveref}
\usepackage{orcidlink}
\usepackage{mathrsfs}
\usepackage{placeins} 

\DeclareMathAlphabet{\mathbbold}{U}{bbold}{m}{n}
\definecolor{AC}{rgb}{1,0.5,0}

\begin{document}

\title{Many-body mobility edges in 1D and 2D revealed by convolutional neural networks}

\author{Anffany Chen\,\orcidlink{0000-0002-0926-5801}}
\email{anffany@ualberta.ca}
\affiliation{Theoretical Physics Institute, University of Alberta, Edmonton, Alberta T6G 2E1, Canada}
\affiliation{Department of Physics, University of Alberta, Edmonton, Alberta T6G 2E1, Canada}

\date{\today}

\begin{abstract}
We adapt a machine-learning approach to study the many-body localization transition in interacting fermionic systems on disordered 1D and 2D lattices. We perform supervised training of convolutional neural networks (CNNs) using labelled many-body wavefunctions at weak and strong disorder. In these limits, the average validation accuracy of the trained CNNs exceeds 99.95\%. We use the disorder-averaged predictions of the CNNs to generate energy-resolved phase diagrams, which exhibit many-body mobility edges. We provide finite-size estimates of the critical disorder strengths at $W_c\sim2.8$ and 9.8 for 1D and 2D systems of 16 sites respectively. Our results agree with the analysis of energy-level statistics and inverse participation ratio. By examining the convolutional layer, we unveil its feature extraction mechanism which highlights the pronounced peaks in localized many-body wavefunctions while rendering delocalized wavefunctions nearly featureless.
\end{abstract}

\maketitle
%%%%%%%%%%%%%%%%%%%%%%%%
%%%    MAIN TEXT     %%%
%%%%%%%%%%%%%%%%%%%%%%%%

\section{Introduction}
Artificial neural networks have proven to be valuable assets in tackling a wide range of problems in condensed matter physics \cite{DasSarma2019,Carleo2019}. With their remarkable ability to discern universal features from extensive datasets and generalize to new, unseen data, neural networks can be trained to perform quantum state tomography \cite{Torlai2018}, accelerate ab initio calculations \cite{Balabin2009, Ryczko2019, Li2022}, and classify various phases of matter based on numerical \cite{Ohtsuki2016, Schindler2017, Carrasquilla2017, vanNieuwenburg2017, Zhang2017, Hu2017, Broecker2017, Chng2017, Liu2018, Doggen2018, Matty2019, Zhang2019, Theveniaut2019, Huembeli2019, Rao2020, Kausar2020, Theveniaut2020} and experimental data \cite{Zhang2019Exp,Rem2019,Bohrdt2019,Ghosh2020}. As universal function approximators \cite{Hornik1989,Hornik1991}, neural networks have also been utilized as variational ansatz for many-body quantum states \cite{Carleo2017,Deng2017,Nomura2017,Choo2018,Sharir2020}, achieving ground-state estimations on par with the state-of-the-art conventional methods. %\cite{Carleo2017,Nomura2017}

One notable application of neural networks is in characterizing the many-body localization transition between an ergodic many-body quantum system, following the eigenstate thermalization hypothesis (ETH) \cite{Deutsch1991,Srednicki1994,Rigol2008}, and a many-body localized (MBL) phase under strong disorder \cite{Basko2006,Nandkishore2015,Abanin2019}. According to ETH,  an isolated, quantum many-body system goes through quantum thermalization over time by acting as its own heat bath, with all local observables eventually assuming thermal expectation values. Introducing sufficiently strong disorder can induce a transition into the MBL phase, where all energy eigenstates become localized, rendering the system non-ergodic and unable to self-thermalize.  Therefore the striking signature of the MBL phase is a partial retention of the initial condition over long times. This phenomenon has been experimentally observed in 1D and 2D ultracold gases \cite{Schreiber2015,Choi2016} and may potentially serve as a mechanism for robust quantum memory. 

To characterize the ETH-MBL transition, the conventional numerical approach is to perform finite-size scaling analyses on simulated data of observables (such as level statistics) over a range of system sizes \cite{Oganesyan2007,Pal2010,Bera2015,Luitz2015,Khemani2017,Panda2019,Sierant2020,Suntajs2020,Abanin2021,Aramthottil2021,Morningstar2022}. This task is computationally demanding due to the exponential growth of Fock space with system size  $N$. Furthermore, the ETH-MBL transition is known to suffer strongly from finite-size effect, such that the apparent phase transition drifts towards strong disorder as $N$ increases. Extrapolating finite-size results to the thermodynamic limit through data collapses is therefore subject to ambiguity, particularly as numerically accessible system sizes are limited to $N\sim\mathcal{O}(10)$ sites \cite{Pietracaprina2018}. Currently, a consensus on the scaling theory for this transition remains elusive. %\cite{Suntajs2020,Aramthottil2021}

Machine learning offers a promising, alternative approach for characterizing the ETH-MBL transition. Recent studies \cite{Schindler2017,Doggen2018,Zhang2019,Theveniaut2019,Huembeli2019,Rao2020,Kausar2020,Theveniaut2020} have successfully automated the classification of the phases using data obtained from exact diagonalization of model Hamiltonians, most of which are 1D spin models. The types of data considered in these studies include many-body energy spectra, many-body wavefunctions, entanglement spectra of these wavefunctions, and other variants obtained through additional feature engineering. A range of learning algorithms, both supervised and unsupervised, have been implemented, including the use of support vector machines and various neural network architectures. Notably, these studies  show that machine learning can provide a finite-size estimate of the phase transition using data from only a single system size. Without the need for scaling analysis to locate the transition point, phase diagrams can be efficiently generated. 

\begin{figure}
\includegraphics[width=0.95\linewidth]{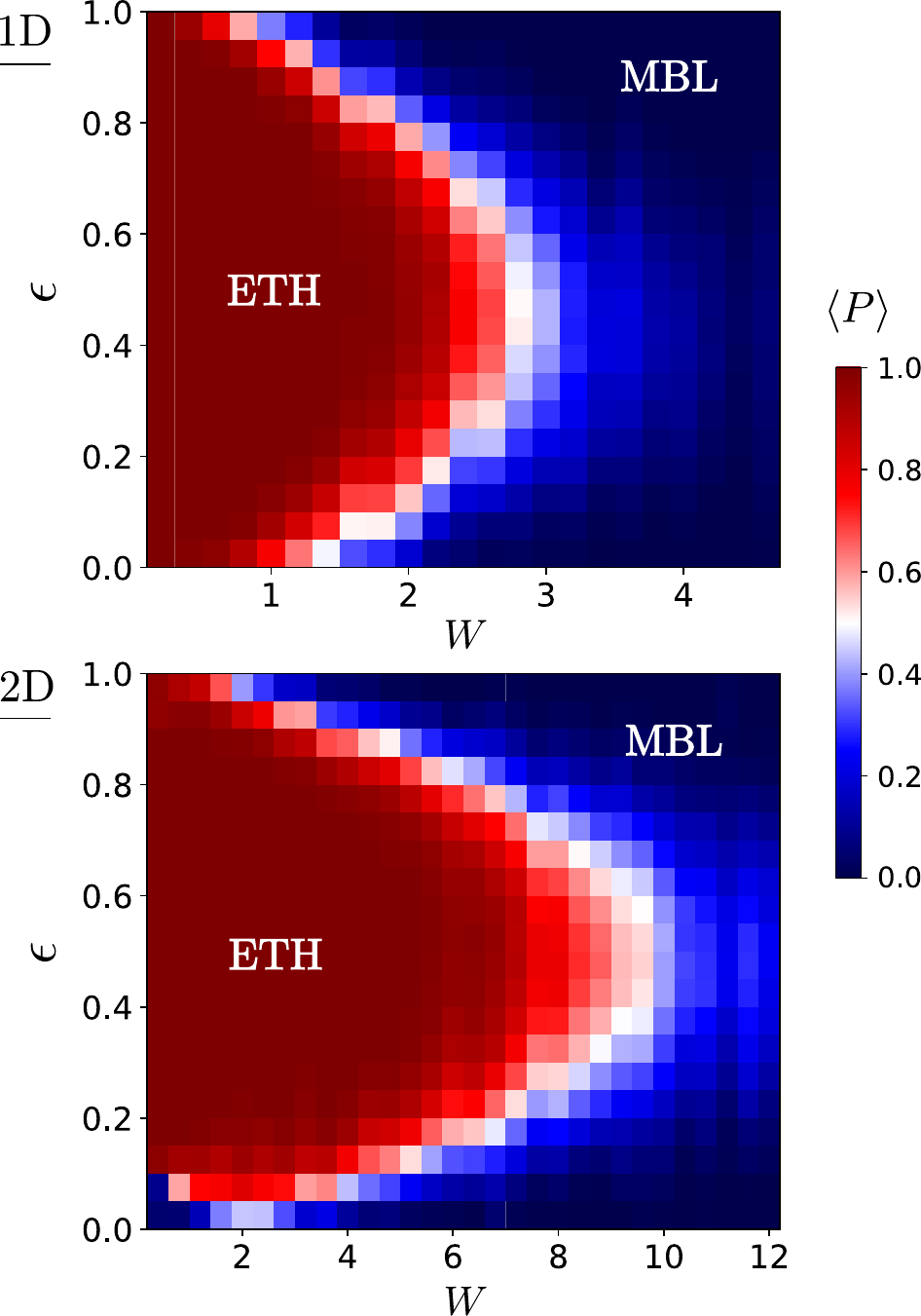}
\caption{\textbf{Machine-predicted phase diagrams.} Trained on many-body wavefunctions at the weak and strong disorder limits, our CNNs effectively generalize to classify wavefunctions near the transition region. We generate the phase diagrams of 16-site 1D and 2D disordered fermionic $t$-$V$ models described by Eq.~\eqref{eq:tV} using the disorder-averaged CNN prediction $\langle P\rangle$, representing the probability of the ETH phase. In both phase diagrams, the many-body mobility edge is clearly visible as the division between ETH (red) and MBL (blue) phases. The finite-size estimates of the critical disorder strengths are $W_c\sim2.8$ in 1D and $W_c\sim9.8$ in 2D.}
\label{fig:nn_pd} 
\end{figure}

In this study, we employ the machine learning approach to investigate the ETH-MBL transition in interacting fermionic systems on 1D and 2D disordered lattices, each consisting of 16 sites. Our method strategically pairs unprocessed many-body wavefunctions as input data with convolutional neural networks (CNNs) for classification. Designed for image recognition, CNNs are expected to be equally suited for learning the local features in many-body wavefunctions. We train CNNs with a simple architecture to differentiate wavefunctions sampled from deep within the ETH and MBL phases, each labeled according to its respective phase. The trained CNNs are then tasked with classifying wavefunctions from the intermediate region. The disorder-averaged predictions of the CNNs are used to construct phase diagrams over energy density $\epsilon$ and disorder strength $W$, which clearly display the many-body mobility edges in 1D and 2D as shown in Fig.~\ref{fig:nn_pd}. 

In the following, we first introduce the fermionic $t$-$V$ model on 1D and 2D disordered lattices (Sec.~\ref{sec:t-V}) and detail our procedure for collecting eigenstate samples via exact diagonalization (Sec.~\ref{sec:ED}). We delve into the supervised training of our neural-network phase classifiers, describing the input data (Sec.~\ref{sec:input}), convolutional network architecture (Sec.~\ref{sec:CNN}), and training techniques (Sec.~\ref{sec:training}).  We then present the energy-resolved phase diagrams based on the predictions of trained CNNs (Sec.~\ref{sec:nn_pd}), compare them with the transition behaviors of energy-level statistics and inverse participation ratio (IPR) (Sec.~\ref{sec:conventional}), and interpret the decision-making mechanism of our trained CNNs (Sec.~\ref{sec:interpret}).

\section{Fermionic $t$-$V$ models with disorder}

\subsection{Model construction\label{sec:t-V}}
We consider repulsive spinless fermions hopping on 1D and 2D lattices with random on-site potentials. The Hamiltonian is given by 
\begin{equation} \begin{split}
H=&\underset{\langle i,j\rangle}{\sum}\left[-t\,(c_{i}^{\dagger}c_{j}+c_{j}^{\dagger}c_{i})+V\left(n_{i}-\frac{1}{2}\right)\left(n_{j}-\frac{1}{2}\right)\right] \\
&+\stackrel[i=1]{N}{\sum}u_{i}\left(n_{i}-\frac{1}{2}\right)\label{eq:tV}
\end{split} \end{equation} 
where $c_{i}^{\dagger}$ creates a spinless fermion at site $i$, $n_i=c_{i}^{\dagger}c_{i}$ is the number operator, $\langle i,j\rangle$ goes over combinations of nearest neighbors, $N$ is the system size, $t$ is the hopping amplitude, $V$ is the strength of the nearest-neighbor repulsive interaction, and $u_{i}$ are on-site potentials randomly drawn from a uniform distribution in the range $[-W,W]$. For $t=1/2$ and $V=1$, Eq.~\eqref{eq:tV} on a 1D chain can be exactly mapped via Jordan-Wigner transformation to a spin-1/2 antiferromagnetic Heisenberg chain subject to a random field in the $z$-direction \cite{affleck1989field,Derzhko2001}:
\begin{equation} 
H=\underset{\langle i,j\rangle}{\sum}\boldsymbol{S}_{i}\cdot\boldsymbol{S}_{j}+\stackrel[i=1]{N}{\sum}u_{i}S_{i}^{z} 
\label{eq:Heisenbert}
\end{equation} 
This 1D spin/fermionic model is well-studied, with its critical disorder estimated to be $W_{c}\sim3.5$ at zero total magnetization, $\sum_{i=1}^{N}S_{i}^{z}=0$, corresponding to the half-filling sector in the fermionic picture \cite{Pal2010,Luitz2015,Khemani2017,Alet2018,Orito2021}. The many-body mobility edge has been demonstrated through finite-size scaling analyses of various observables \cite{Luitz2015,Orito2021} and machine-learning technique \cite{Schindler2017}, which we will use to benchmark our phase diagram in the 1D case. From here onward, we set $t=1/2$ and $V=1$, and focus on half-filling for both the 1D and 2D systems.

The lattice geometries considered here are 1D chain and 2D square lattices, both given periodic boundary conditions to prevent localization by the boundaries. Each lattice consists of $N=16$ sites, with the 2D lattice arranged as $4\times4$. To construct the many-body Hamiltonians, we start by defining the creation operators in the occupancy number basis of the $2^{16}$-dimensional Fock space. The basis states are ordered by the total number of particles $N_f$, such that a particle-number conserving term like $c_i^{\dagger}c_j$ would be block-diagonal with each block corresponding to a specific $N_f$. Using these creation operators, we construct the many-body Hamiltonian as per Eq.~\eqref{eq:tV} and the specified lattice geometries. In the following analysis, we focus on the half-filling sector, $N_f=8$, described by the $\mathcal{D}\times \mathcal{D}$ diagonal block with $\mathcal{D}=N\text{ choose }N_f=12870$.

\subsection{Exact diagonalization \label{sec:ED}}
For training neural networks to differentiate ETH and MBL wavefunctions, we select representative disorder strengths deep within each phase: $(W_{\text{ETH}},W_{\text{MBL}})=(0.2,12)$ for 1D and $(0.4,24)$ for 2D. Choosing these values does not require knowledge of the critical disorder $W_{c}$, because one can always verify that conventional observables follow the expected ETH/MBL behaviors at these values. For the phase diagrams, we define suitable grids of $W$ values in the intermediate regions: $W\in[0.2,4.6]$ for 1D and $[0.4,12]$ for 2D. At each selected $W$ value, we implement 50 random disorder realizations and perform exact diagonalization of the Hamiltonians. Each energy spectrum is normalized as $\epsilon=(E-E_{min})/(E_{max}-E_{min})$ where $\epsilon$ is the energy density and $E_{min}$/$E_{max}$ is the smallest/largest eigenvalue of the spectrum. For the phase diagram in the 2D case, we use 50 additional disorder realizations for each $W$ value in the range $[8.4,10]$.

Eigenstates from each disorder realization are binned into 20 equal energy intervals between $\epsilon=0$ and $1$. In each bin, we discard all but the 50 eigenstates with energy densities closest to the bin's center, greatly reducing data storage and computational demands during subsequent analysis. We do however keep all the eigenvalues for computing the energy-level statistics later. We observe that using such a small sample of eigenstates does not significantly affect the disorder-averaged values of IPR and machine predictions. Note that due to the low density-of-states near $\epsilon=0$ and $1$, bins in these regions contain fewer than 50 eigenstates per disorder realization.
%Note that while full ED is sufficiently efficient for $N=16$, it may be more efficient use the shift-invert method for larger system sizes, in which a fixed number of eigenpairs per energy bin is computed. 

\section{Neural-Network Phase Classifier \label{sec:method}}
Neural networks are complex, nonlinear functions consisting of alternating layers of linear and nonlinear maps. The linear maps are defined by a large number of adjustable parameters,  \textit{weights} and \textit{biases}. The nonlinear maps are \textit{activation functions} that mimic the behavior of biological neurons, which produce an output only when the input exceeds a certain threshold. A given neural network  can be trained to approximate any function to a certain degree of accuracy. In \textit{supervised learning}, the model is provided with a training dataset consisting of input-output pairs, and the weights and biases are  adjusted to minimize a  \textit{loss function}, which quantifies the difference between the model's predictions and the correct outputs. We refer to Refs.~\onlinecite{Nielsen2015Neural,Neupert2022Introduction} for comprehensive introduction to neural networks and machine learning.

Our objective is to train a neural network to approximate the hypothetical function which maps a many-body wavefunction to the correct binary classification, ETH (labelled 1) or MBL (labelled 0). Our neural network would merely be an approximation to this function, so its output would not be binary but rather a continuous real number $P$ ranging from 0 to 1, representing the  probability that the input wavefunction belongs to the ETH phase. Upon disorder averaging, the model's prediction can be regarded as an order parameter $\langle P\rangle$, transitioning from 1 in the ETH phase to 0 in the MBL phase. Assuming that the trained model is not biased towards one phase over the other,  the point  $\langle P\rangle=0.5$ can be interpreted as the critical point for the $N=16$ systems considered here. %In contrast, conventional observables require finite-size scaling analyses over various system sizes to locate the transition.

\subsection{Input data \label{sec:input}}
For the input data, we use the probability densities $|\Psi_{j}|^{2}$ of the many-body wavefunctions, where $j$ labels the the occupancy number basis. This choice is economical because (i) each wavefunction serves as a data sample, unlike using the energy spectrum as input which requires one exact diagonalization per sample, and (ii) it avoids additional feature engineering, such as calculating the entanglement spectra of the wavefunctions, which increases the computational costs. Moreover, this approach does not assume a priori wavefunction behaviors in either phase, leading to data-driven results. %Input data based on the eigenstates, rather than the eigenvalues, requires fewer exact diagonalizations, as each eigenstate gives one data sample. Eigenstates can be used directly as input or further engineered by, for example, deriving their entanglement spectra. While the entanglement spectra reduce the input lengths, it does requires more matrix diagonalizations and is only meaningful in systems with open boundary conditions \cite{Schindler2017}. 

Every time we train a model, we prepare a set of labelled data by randomly selecting 10,000 ETH and 10,000 MBL wavefunctions with energy densities $0.15<\epsilon<0.85$ collected at disorder strengths $W_{\text{ETH}}$ and $W_{\text{MBL}}$ as discussed in Sec.~\ref{sec:ED}, and pair them with outputs of 1 and 0 respectively. We observe that the MBL wavefunctions are typically localized on a subset of basis states, displaying a few highly pronounced peaks, while the ETH wavefunctions are distributed across all basis states. This visible difference motivates our choice of using CNNs to classify the wavefunctions, as CNNs are designed for image recognition. Our input probability densities can be viewed as 1 by 12870 grayscale images, with each pixel representing the probability at a specific basis state. 

\subsection{Convolutional neural network \label{sec:CNN}}

\begin{figure}
\includegraphics[width=0.9\linewidth]{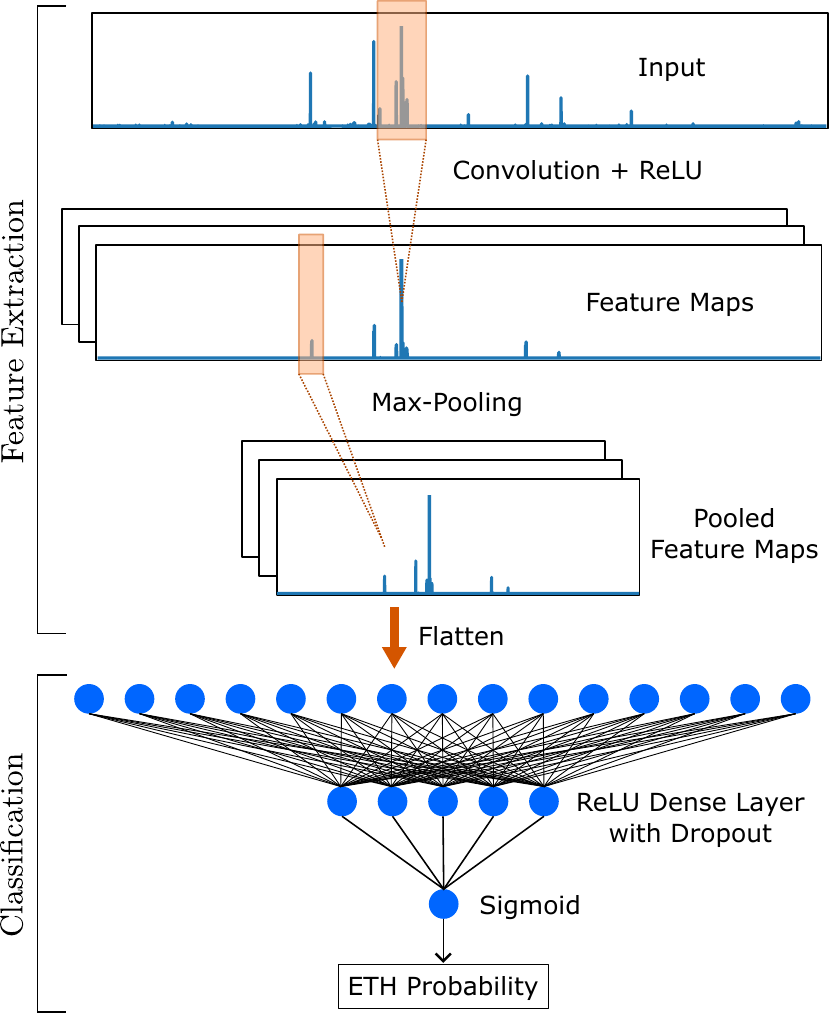}
\caption{\textbf{Convolutional neural network architecture.} For the task of classifying many-body wavefunctions, we design a simple CNN composed of (i) a convolutional layer followed by max-pooling for feature extraction, (ii) a dense layer with dropout regularization for classification, and (iii) an output layer consisting of a single sigmoid neuron for predicting the probability for the input wavefunction to be in the ETH phase. All activation functions are chosen to be ReLU except in the output layer.}
\label{fig:cnn} 
\end{figure}

As shown in  Fig.~\ref{fig:cnn}, our simple CNN consists of  (i) a convolutional layer followed by a max-pooling layer for feature extraction,  (ii) a dense layer with dropout regularization for classification, and (iii) an output layer of one sigmoid neuron for prediction. These layers are built in Python using the TensorFlow package \cite{tensorflow}. In the following we describe the operations in each layer and their purposes.

\emph{Convolutional layer.---}The linear map in this layer is the convolution operation between  \textit{kernels} (or \textit{filters}) and  the input data. The total number of kernels, $m$, and the length of each kernel, $l$, are \textit{hyperparameters} which are fixed prior to the training process. Having multiple kernels are crucial for detecting different local features in the input data. Each kernel slides across the input with a stride of 1, computing dot products between its array of  $l$ weights and the corresponding segment of the input data it covers at each position. Each dot product,  added with the kernel's bias, is passed through a nonlinear ReLU (Rectified Linear Unit) activation function defined as
\begin{equation}
    \text{ReLU}(x)=\max(0,x)
\end{equation}
which sets negative values to zero. The outputs of the ReLU neurons form a feature map, so  $m$ kernels give rise to  $m$ 
 feature maps. The weights and biases in the kernels are initialized randomly and optimized during the training process. 

\emph{Max-pooling layer.---}This layer performs a down-sampling operation dictated by a hyperparameter $p$. It slides a window of size $p$ (with a stride of  $p$) across each feature map and selects the maximum value within each window. This process results in $m$ pooled feature maps with length reduced by a factor of $p$.

\emph{Dense layer.---}The pooled feature maps are flattened into a single 1D vector $v$ of length $L$, which is then fed into a dense (or fully connected) layer. A hyperparameter $q$ dictates the number of ReLU neurons in this layer. The linear operation  \begin{equation}
    f(v)=Av+b,
\end{equation}
where $A$ is the $q\times L$  weight matrix and $b$  is the bias vector, maps $v$ to a vector of length $q$.  The ReLU activation function is then applied to the resulting vector element-wise. The weights and biases in $A$ and $b$ are optimized during training.  

\emph{Dropout layer.---}A dropout layer following the dense layer randomly deactivates a fraction $d$ of the neurons in the dense layer during training by setting their activation functions to zero. This is a regularization technique designed to prevent \textit{overfitting} to non-universal features specific to the training data. The introduced randomness prevents any single neuron in the dense layer from becoming too specialized to certain patterns from the training data, encouraging the model to learn more robust and universal feature. The dropout layer only operates during training.

\emph{Output layer.---}In the final layer, the output of the dense layer is multiplied by a $1\times q$ weight matrix and then combined with a bias; as before, the weights and bias are optimized during training. The resulting value is passed through a sigmoid activation function given by 
\begin{equation}
    \sigma(x)=\frac{1}{1+e^{-x}},
\end{equation}
which  maps any real number into a value $P\in[0,1]$. For the task of binary classification, $P$ can be interpreted as the probability for the "1" class. In this case, it represents the probability for the input wavefunction to be in the ETH phase.

%Given that our input arrays are longer than those in studies employing feature engineering, like truncation \cite{Theveniaut2019} or entanglement spectra \cite{Schindler2017}, our use of an initial convolutional layer circumvents the potential issue of excessive parameters. The convolutional layer applies the same filter – a small matrix of weights – across the entire input array, efficiently extracts local features irrespective of the input lengths.

\subsection{Supervised training \label{sec:training}}
\emph{Loss function.---}Supervised training of our CNN amounts to tuning the weights and biases to minimize the difference between model predictions and the correct labels provided in the training dataset. Here the difference is measured by the \textit{binary cross entropy}, which is a common choice of loss function for binary classification:
\begin{equation} 
\text{Loss}=-(y\log(P)+(1-y)\log(1-P)) \label{eq:loss} 
\end{equation} where $y$ is the correct label and $P$ is the model prediction. When the label is 0, the first term in Eq.~\eqref{eq:loss} vanishes and the loss function is approximately $P$ for $P\sim0$. Similarly, when the label is 1, the loss function approximately measures how far $P$ strays from 1. 

\emph{Gradient descent.---}The dataset prepared in Sec.~\ref{sec:input} is randomly split 50/50 into a training set and a test set for evaluating the model's performance on unseen data. The training process is organized into \textit{epochs}, each is a complete pass through the entire training set. During each epoch, the training set is randomly divided into smaller \textit{batches} for \textit{mini-batch gradient descent}. We typically set the batch size to be 50 samples. Each batch goes through the model's layers, a process known as \textit{feedforward}. For each sample in the batch, a loss is computed according to Eq.~\eqref{eq:loss}. Then the \textit{backpropagation} algorithm calculates the gradient of the loss  with respect to the model parameters. This gradient is averaged over all samples in the batch. The optimization algorithm uses this average gradient to update the parameters, moving them in the direction of the steepest descent (opposite to the gradient), with the magnitude of change controlled by the \textit{learning rate}. As opposed to fixing the same learning rate for all parameters, we have opted to use TensorFlow's Adam optimizer, which adjusts the learning rate for each parameter throughout the training process.

\emph{Training history.---}At the end of each epoch, the average loss over the test set is computed and recorded, known as the validation loss. A successful training is marked by a decreasing trend in both training loss (average loss over training data) and validation loss as training progresses over successive epochs, which indicates that the model is learning and generalizing well to unseen data. However, an increase in validation loss coupled with a decrease in training loss can signal potential overfitting, where the model memorizes the training data instead of learning generalizable features. To prevent significant overfitting, we implement \textit{early stopping}, which halts training when validation loss stops decreasing for a few consecutive epochs.

\emph{Cross validation.---}Due to the inherent randomness in the training process, including weight/bias initialization and data sampling, model training can vary with each iteration, often converging to different local minima in the loss function landscape. To evaluate model performance reliably, we perform $k$-fold cross validation, training the model $k=20$ times with different training and test sets. Examining all training histories together enables accurate assessment of the model's performance under a given set of hyperparameters, which facilitates hyperparameter tuning.

\emph{Hyperparameters.---}Through experimentation, we determined that the following set of hyperparameters yields optimal validation loss at the end of training --- $m=16$, $l=10$, $p=2$, $q=60$, and $d = 0.2$ --- for models trained with wavefunctions of the 1D system. This set of hyperparameters leads to 99.99\% validation accuracy (averaged over $k$ folds of training). Validation accuracy is defined as the percentage of correct predictions on the test data, considering a prediction correct if its rounded integer value matches the label. Minor variations in these hyperparameters do not significantly affect performance. We thus fix the hyperparameters at these values when training CNN models for the 1D system.

However, we find that this set of hyperparameters is suboptimal for training with wavefunctions of the 2D system, resulting in an average test accuracy of 99.74\%. Further experimentation reveals that increasing the kernel size $l$ from 10 to 100 and the dropout rate $d$ from 0.2 to 0.5 significantly improves test accuracy, achieving an average of 99.97\%. The improved performance due to a larger kernel suggests that the important local features in wavefunctions of the 2D system likely span a wider range of basis states. Accordingly, we fix the hyperparameters at $m=16$, $l=100$, $p=2$, $q=60$, and $d = 0.5$ for the 2D case.

\section{Results and Discussion}

\subsection{Machine-predicted phase diagrams \label{sec:nn_pd}}
To generate the energy-resolved phase diagram of the 1D fermionic chain, we trained 20 CNN models, with architecture described in Sec.~\ref{sec:CNN} and hyperparameters in Sec.~\ref{sec:training}, using wavefunctions at disorder strengths $W_{\text{ETH}} = 0.2$ and $W_{\text{MBL}} = 12$ (see Sec.~\ref{sec:ED} and \ref{sec:input} for details on training data). Similarly in the 2D case, we trained another 20 CNN models, this time using wavefunctions of the 2D fermionic system at $W_{\text{ETH}} = 0.4$ and $W_{\text{MBL}} = 24$. Our trained models demonstrate over 99.95\% validation accuracy in predicting the correct phases of the test wavefunctions.

Exploiting neural networks' generalization capacity, we input wavefunctions from the intermediate regions. At every pair of discretized energy density $\epsilon$ and disorder strength $W$, we first averaged one trained CNN's prediction over wavefunctions belonging to one disorder realization. These averages were then further averaged over all disorder realizations and 20 CNNs. The disorder- and training-averaged probability for the ETH phase forms the phase diagrams shown in Fig.~\ref{fig:nn_pd}. Note that the models trained on wavefunctions from the 1D and 2D systems were specifically used to produce their respective 1D and 2D phase diagrams. We further note that we did not follow the common practice of truncating the phase diagrams at low and high energies, since the average prediction of our CNNs appears convergent despite the scarcity of data in these regions.

In both phase diagrams, the mobility edge is clearly visible as the division between ETH (red) and MBL (blue) phases. For the 1D system, the mobility edge agrees with previous studies \cite{Luitz2015,Schindler2017}, exhibiting a characteristic bell shape with the tip dropping slightly below $\epsilon=0.5$. We estimate the critical disorder to be $W_c\sim2.8$, agreeing with a previous machine-predicted estimate for the $N=16$ chain \cite{Schindler2017}. Our finite-size estimate of $W_c$ is smaller than the thermodynamic limit $W_c\sim3.5$ determined through finite-size scaling analyses \cite{Pal2010,Luitz2015,Khemani2017,Alet2018,Orito2021}. This difference is expected due to the strong finite-size effect at the ETH-MBL transition. 

In comparison, the phase diagram of the 2D system exhibits notable differences. The tip of the mobility edge is more aligned to $\epsilon=0.5$. Moreover, the eigenstates at small $W$ are predicted to have high probability of localization near $\epsilon=0$, in contrast to the 1D case where no states are localized at small $W$. The estimated critical disorder is $W_c\sim9.8$, significantly greater than the 1D case. The increase in critical disorder with higher spatial dimension is a well-known phenomenon in the non-interacting limit, and can be understood in terms of classical random walks on lattices \cite{chen2023anderson}.

\subsection{Comparison with conventional observables\label{sec:conventional}} 

\begin{figure}
\includegraphics[width=\linewidth]{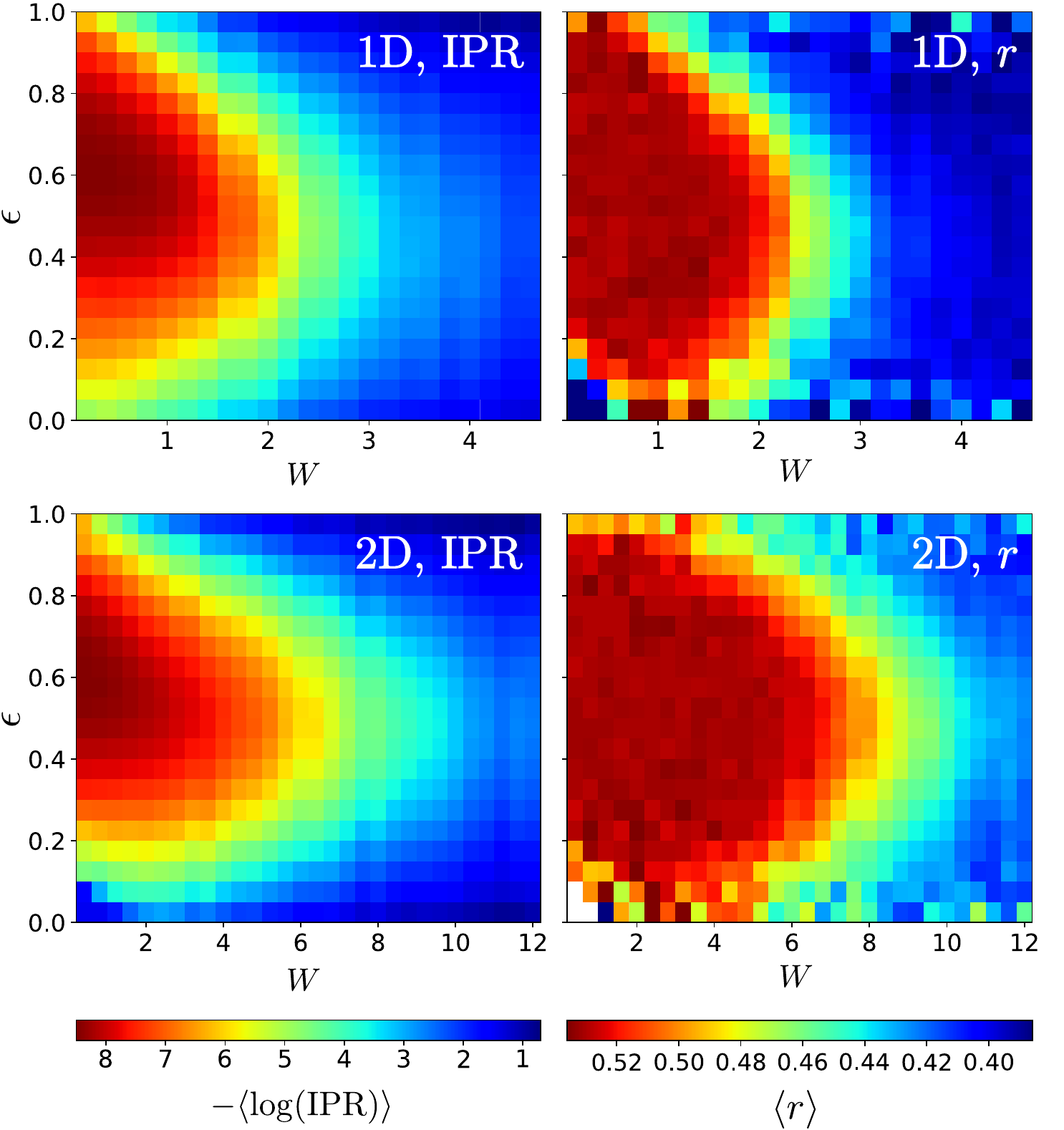}
\caption{\textbf{Inverse participation ratio and energy-level statistics.} For the 16-site 1D and 2D systems, disorder-averaged $\langle\log(\text{IPR})\rangle$ and gap ratio $\langle r\rangle$ as functions of energy density $\epsilon$ and disorder strength $W$ bear similar qualitative features as the machine-predicted phase diagrams (Fig.~\ref{fig:nn_pd}). For $\langle r\rangle$, the max and min values of the colorbar correspond to $\langle r\rangle_{\text{GOE}}$ and $\langle r\rangle_{\text{P}}$ respectively, and the noise is due to insufficient eigenvalue data near $\epsilon=0$ and $1$.}
\label{fig:conventional_pd} 
\end{figure}

To verify the machine-predicted phase diagrams, we analyze two conventional observables across the transition. The first observable is the energy-level statistics based on the gap ratios in the many-body energy spectrum ~\cite{Oganesyan2007}: 
\begin{equation} 
r_{\alpha}=\frac{\min\{\epsilon_{\alpha+1}-\epsilon_{\alpha},\epsilon_{\alpha}-\epsilon_{\alpha-1}\}}{\max\{\epsilon_{\alpha+1}-\epsilon_{\alpha},\epsilon_{\alpha}-\epsilon_{\alpha-1}\}} 
\end{equation} where $\epsilon_{\alpha}$ are the sorted energy densities of a single disorder realization. Using the full energy spectrum of each disorder realization, we compute the averaged gap ratio within each of the 20 energy intervals between $\epsilon=0$ and 1. These values are then averaged over all disorder realizations. In the ETH phase, the expected value of the disorder-averaged gap ratio $\langle r\rangle$ is $\langle r\rangle_{\text{GOE}}=4-2\sqrt{3}=0.536$~\cite{Atas2013}, corresponding to the Wigner surmise of the Gaussian orthogonal ensemble (GOE). In the MBL phase, the level statistics is described by the Poisson distribution, averaging to $\langle r\rangle_{\text{P}}=2\ln2-1=0.386$.

The second observable is the inverse participation ratio (IPR), defined as \begin{equation} \text{IPR}(\Psi)=\stackrel[j=1]{\mathcal{D}}{\sum}|\Psi_{j}|^{4}, \end{equation} where $\Psi$ is a many-body wavefunction and $j$ goes over the occupancy number basis in the half-filling sector. The inverse of IPR quantifies the support of $\Psi$ in our choice of basis. For each disorder realization, we average the IPR of up to 50 eigenstates per energy interval (see Sec.~\ref{sec:ED} for details on data sampling). These values are then averaged over all disorder realizations. In contrast to the Anderson localization transition, where $\langle\text{IPR}\rangle$ defined in the real-space position basis increases from $1/N$ to $1$ towards the strong disorder limit, the $\langle\text{IPR}\rangle$ in the Fock-space basis increases more gradually. Unlike the former, it does not approach $1$ because localized many-body wavefunctions have nonzero support over many basis states even at strong disorder. Thus we compute the disorder-average of log(IPR) to highlight the transition from ETH to MBL.

In Fig.~\ref{fig:conventional_pd}, we plot $\langle r\rangle$ and $\langle\log(\text{IPR})\rangle$ as functions of $\epsilon$ and $W$. For these conventional observables, pinpointing the transition boundary involves analyses over various system sizes and finite-size scaling. Thus we focus on the contours, which are lines of equal value. In both the 1D and 2D cases, the $\langle \log(\text{IPR})\rangle$ contour at around $-3.5$ is very similar to the mobility edge in Fig.~\ref{fig:nn_pd}. On the other hand, the contours of $\langle r\rangle$, while consistent with the mobility edges, have less "pointy" profiles with weaker curvature. Note that our data obtained from 50 disorder realizations are insufficient for converging $\langle r\rangle$ near $\epsilon=0$ and $1$ where eigenvalues are scarce.

\subsection{Model interpretation\label{sec:interpret}}

\begin{figure*}
\includegraphics[width=0.9\linewidth]{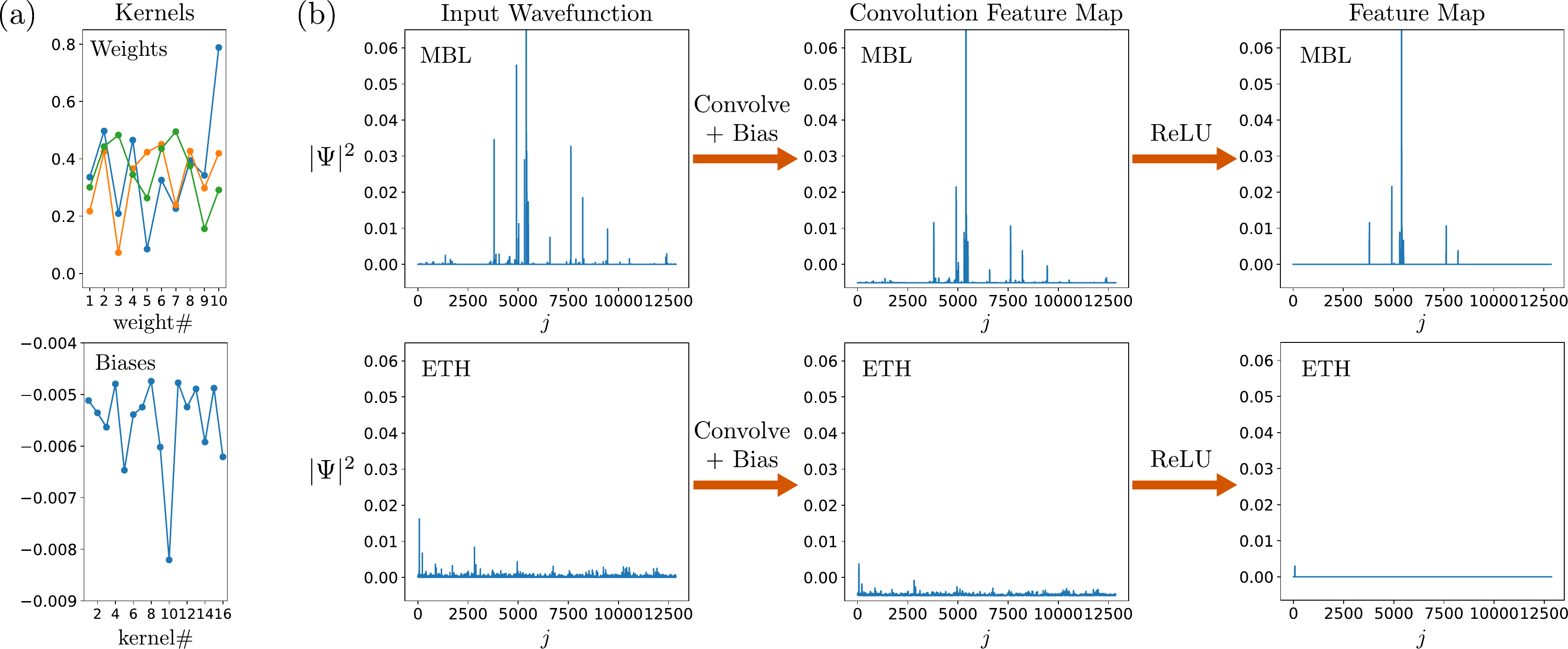}
\caption{\textbf{Interpretation of the convolutional layer.} (a, top panel) The kernel weights in our trained CNNs are highly fluctuating values between -1 and 1. Here we show the weights of three kernels (each with $l=10$ weights) belonging to a CNN trained on the wavefunctions of the 1D system. (a, bottom panel) The kernel biases (one bias per kernel) of the same CNN are small negative numbers. (b) %The key difference between ETH and MBL probability densities $|\Psi|^2$ is highlighted by the convolutional layer. 
By learning to apply a small negative bias right before ReLU activation, a typical kernel in our trained CNNs truncates small values in the input probability density $|\Psi|^2$, which renders an ETH wavefunction nearly featureless and accentuates the pronounced profile of a MBL wavefunction. This feature extraction mechanism, demonstrated here with wavefunctions of the 1D system, is observed in both CNNs trained on 1D and 2D systems.}
\label{fig:feature} 
\end{figure*}

To understand the decision-making of our trained CNNs, we examine the kernel weights and biases and the feature maps generated from the input data. Prior to classification by the dense layer, the input probability densities $|\Psi|^2$  undergo a series of operations: convolution with the kernels plus biases (resulting in convolution feature maps), ReLU activation (feature maps), and max-pooling (pooled feature maps). For both CNNs trained on the 1D and 2D systems, the kernels generally have highly fluctuating weights between -1 and 1, along with small negative biases (see Fig.~\ref{fig:feature}(a) for examples in the 1D case). During convolution, these weights effectively scale down  $|\Psi|^2$,  which is then shifted downward by the negative biases. The negative values in the resulting convolution feature maps, due to the negative biases, are truncated by the ReLU activation. Lastly, the max-pooling layer, with a minimal pool size of 2,  downsamples the feature maps by a factor of two without significantly altering the extracted features.

Fig.~\ref{fig:feature}(b) shows the feature extraction process by a typical kernel applied to an ETH/MBL wavefunction of the 1D system. The ETH wavefunction, characterized by its low probability density at almost all basis states, becomes nearly featureless after the application of the negative bias followed by ReLU activation. In contrast, the same operation accentuates the pronounced peaks in the MBL wavefunction, reducing smaller signals to zero while preserving the more significant ones. This explains how our convolutional layer effectively highlights the key differences between MBL and ETH wavefunctions, thereby simplifying the classification task for the dense layer.

\section{Conclusion}
In this work, we investigated interacting spinless fermionic systems on 1D and 2D lattices with random on-site potentials, focusing on systems of 16 sites. Through exact diagonalization, we collected many-body wavefunctions from various disorder realizations. We then conducted supervised training of neural networks using wavefunctions at weak and strong disorder, labeled as ETH and MBL respectively. We specifically chose convolutional neural networks (CNNs) for their capability in local pattern recognition. Utilizing effective training techniques, including dropout regularization and cross validation, our CNNs achieved over 99.95\% accuracy on test data, successfully classifying wavefunctions deep in the ETH and MBL phases.

Leveraging the neural network's generalization ability, we provided the CNNs with wavefunctions near the transition region and used the disorder-averaged prediction $\langle P\rangle$, representing the probability for the ETH phase, to construct phase diagrams. The energy-resolved phase diagrams over energy density $\epsilon$ and disorder strength $W$ precisely locate the many-body mobility edges in both 1D and 2D systems. We estimated the critical disorder strengths to be $W_c\sim2.8$ for 1D and $W_c\sim9.8$ for 2D, applicable to finite-sized systems of 16 sites.

Our analysis of energy-level statistics and inverse participation ratio corroborates our phase diagrams by showing similar qualitative features. We further examined the CNN's weights, biases, and feature maps, gaining insights into its feature extraction mechanism. We found that the convolutional layer has learned to truncate small values in the input probability densities through negative biases and ReLU activation, effectively retaining only the strong input signals for classification. This mechanism was observed in both CNNs trained on 1D and 2D systems, demonstrating its applicability across different dimensions. Future studies could investigate its connection to conventional observables or potentially formulate new order parameters inspired by the learned mechanism.

The ultimate success of the machine-learning approach for characterizing the ETH-MBL phase boundary hinges on precise quantification of the machine's predictions. This includes quantifying the uncertainties in the predictions and conducting finite-size scaling analysis to extend finite-size results to the thermodynamic limit. Assessing the effectiveness of \textit{transfer learning}, particularly by applying CNNs trained on 1D systems to classify wavefunctions in 2D systems and vice versa, could reveal whether the machine-based order parameter $P$ is universal, independent of lattice configurations and spatial dimensions. Lastly, one could experiment with neural networks with more advanced architectures, which may generalize better to the transition region and lead to more precise determination of the phase boundary.
 
\emph{Acknowledgements.---}This research benefited greatly from discussions with Igor Boettcher, Joseph Maciejko, Canon Sun, Santanu Dey, and Davidson Noby Joseph. The numerical computation was enabled in part by support provided by Compute Ontario (\href{https://www.computeontario.ca/}{computeontario.ca}) and the Digital Research Alliance of Canada (\href{https://alliancecan.ca}{alliancecan.ca}). The author gratefully acknowledges the support of NSERC Discovery Grant RGPIN-2020-06999, Avadh Bhatia Fellowship, startup fund UOFAB Startup Boettcher, and the Faculty of Science at the University of Alberta.

%\bibliographystyle{apsrev4-1}
%\bibliography{biblio}

\let\oldaddcontentsline\addcontentsline
\renewcommand{\addcontentsline}[3]{}
\bibliography{mbl_biblio}
\let\addcontentsline\oldaddcontentsline 

\cleardoublepage

\setcounter{equation}{0}
\setcounter{figure}{0}
\setcounter{table}{0}
\renewcommand{\theequation}{A\arabic{equation}}
\renewcommand{\thefigure}{A\arabic{figure}}
\renewcommand{\thetable}{A\arabic{table}}

% Reset hyperlinks
\renewcommand\theHtable{Appendix.\thetable}
\renewcommand\theHfigure{Appendix.\thefigure}
\renewcommand\theHequation{Appendix.\theequation}

% Begin Appendix
\setcounter{table}{0}
\renewcommand{\thetable}{A\arabic{table}}
\renewcommand\theHtable{Appendix.\thetable}
\setcounter{figure}{0}
\renewcommand{\thefigure}{A\arabic{figure}}
\renewcommand\theHfigure{Appendix.\thefigure}

%\appendix

%\section{A1. ...}...

%\FloatBarrier

\end{document}